*Chapter 2* **PREPRINT / DRAFT**

# DESIGN AND IMPLEMENTATION OF A SEMANTIC DIALOGUE SYSTEM FOR RADIOLOGISTS

***Daniel Sonntag[1], Martin Huber[2], Manuel Möller[3], Alassane Ndiaye[4], Sonja Zillner[5] and Alexander Cavallaro[6]***

[1,4]DFKI - German Research Center for Artificial Intelligence, Stuhlsatzenhausweg 3, D-66123 Saarbrücken, Germany.
[2]Siemens AG, Corporate Technology, CT SE 5, Günther-Scharowsky-Str. 1, D-91058 Erlangen, Germany.
[3]DFKI - German Research Center for Artificial Intelligence, Trippstadter Straße 122, D-67663 Kaiserslautern, Germany.
[5]Siemens AG, Corporate Technology, CT IC 1, Otto-Hahn-Ring 6, D-81739 München, Germany.
[6]Friedrich-Alexander-Universität Erlangen-Nürnberg, Maximiliansplatz 1, D-91054 Erlangen, Germany.

1 Corresponding author:
E-mail: sonntag@dfki.de, phone: +49 681 857755254; fax: +49 681 857755021.
2 E-mail: martin.huber@siemens.com, phone: +49 9131 735350; fax: +49 9131 733190.
3 E-mail: manuel.moeller@dfki.de, phone : +49 631 20575132; fax: +49 631 20575102.
4 E-mail: ndiaye@dfki.de, phone: +49 681 3025396; fax: +49 681 3025020.
5 E-mail: sonja.zillner@siemens.com, phone: +49 89 63647132; fax: +49 89 63649438.
6 E-mail: alexander.cavallaro@uk-erlangen.de, phone: +49 9131 8545515.

**Keywords**: User/Machine Dialogue, Semantic Data Model, Ontologies, Multimedia Database, Clinical Information System, Health Care Infrastructure, Medical Imaging Systems

## ABSTRACT

This chapter describes a semantic dialogue system for radiologists in a comprehensive case study within the large-scale MEDICO project. MEDICO addresses the need for advanced semantic technologies in the search for medical image and patient data. The objectives are, first, to enable a seamless integration of medical images and different user applications by providing direct access to image semantics, and second, to design and implement a multimodal dialogue shell for the radiologist. Speech-based semantic image retrieval and annotation of medical images should provide the basis for help in clinical decision support and computer aided diagnosis.

We will describe the clinical workflow and interaction requirements and focus on the design and implementation of a multimodal user interface for patient/image search or annotation and its implementation while using a speech-based dialogue shell. Ontology modeling provides the backbone for knowledge representation in the dialogue shell and the specific medical application domain; ontology structures are the communication basis of our combined semantic search and retrieval architecture which includes the MEDICO server, the triple store, the semantic search API, the medical visualization toolkit MITK, and the speech-based dialogue shell, amongst others. We will focus on usability aspects of multimodal applications, our storyboard and the implemented speech and touchscreen interaction design.

## 1. INTRODUCTION

Clinical care and research increasingly rely on digitized patient information. There is a growing need to store and organize all patient data, including health records, laboratory reports and medical images. Effective retrieval of images builds on the semantic annotation of image contents. At the same time it is crucial that clinicians have access to a coherent view of these data within their particular diagnosis or treatment context. This means that with traditional user interfaces, users may browse or explore visualized patient data, but little or no help is given when it comes to the interpretation of what is being displayed. Semantic annotations should provide the necessary image information and a semantic dialogue shell should be used to ask questions about the image annotations while engaging the clinician in a natural speech dialogue at the same time.

Our research activities in the Core Technology Cluster-WP4 (which provides a semantic dialogue shell) are in the context of the MEDICO[1] project. MEDICO addresses the need for advanced semantic technologies in the search for medical image and patient data. It aims for the automatic extraction of meaning from medical images and the seamless integration of the extracted knowledge into medical processes, such as clinical decision making. In other words, the computer will, first, automatically learn to interpret images to catalogue them, second, accurately find them in databases, and third, detect similarities.

A wide range of different imaging technologies in various *modalities* exist, such as 4D 64-slice Computer Tomography (CT), whole-body Magnet Resonance Imaging (MRI), 4D Ultrasound, and the fusion of Positron Emission Tomography and CT (PET/CT). Today, medical images have become indispensable for detecting and differentiating pathologies, planning interventions, and monitoring treatments. While medical images provide a wealth of information to clinicians, current medical image databases, called PACS (Picture Archiving and Communications System), as well as associated Radiology Information Systems (RIS) are still indexed by keywords assigned by humans or indexed by metadata originating from the image acquisition and not the image (region) contents. This limitation severely hampers clinical workflows.

Over the last ten years, the limitations of keyword-based manual image annotation for retrieval motivated the development of content-based image retrieval (CBIR) systems. In these systems, image retrieval additionally includes low-level features, such as color, shape, and texture, which are automatically extracted from the images themselves. However, such CBIR systems face the semantic gap, defined in (Smeulders et al., 2000) as “the lack of coincidence between the information that one can extract from the visual data and the interpretation that the same data have for a user in a given situation.” While not abandoning the strengths of classical CBIR based on comparing low level features for retrieval of, e.g., similar liver lesions, the primary goal of the MEDICO project is retrieval based on *semantic image annotations*.

[1] This research has been supported in part by the THESEUS Program in the MEDICO project, which is funded by the German Federal Ministry of Economics and Technology under grant number 01MQ07016. The responsibility for this publication lies with the authors.

The objective of the Core Technology Cluster-WP4 is to build the next generation of intelligent, scalable, and user-friendly *semantic search interfaces* for the medical imaging domain, based on semantic technologies. Ontology-based knowledge representation is used not only for the image contents, but also for the complex natural language understanding and dialogue management process. With the incorporation of higher level knowledge represented in ontologies, different semantic views of the same medical images (such as aspects of structure, function, and disease) can be explicitly stated and integrated.

We will provide an outline of the design phase, including the discussion of clinical requirements and an overview of our implementations of these requirements. We build upon the developments and implementations of the first phase (2008-2009) to achieve the objectives of the Core Technology Cluster-WP4 and MEDICO and we focus on the challenges, requirements, and possible solutions related to new multimodal interaction metaphors where the information access based on natural speech plays the major role. For more information, visit http://theseus-programm.de/scenarios/en/medico. In this book chapter, we describe the semantic dialogue-based multi-touch installation, i.e., the design and implementation of a semantic dialogue system for radiologists, for improving today’s clinical reporting process (Figure 1).

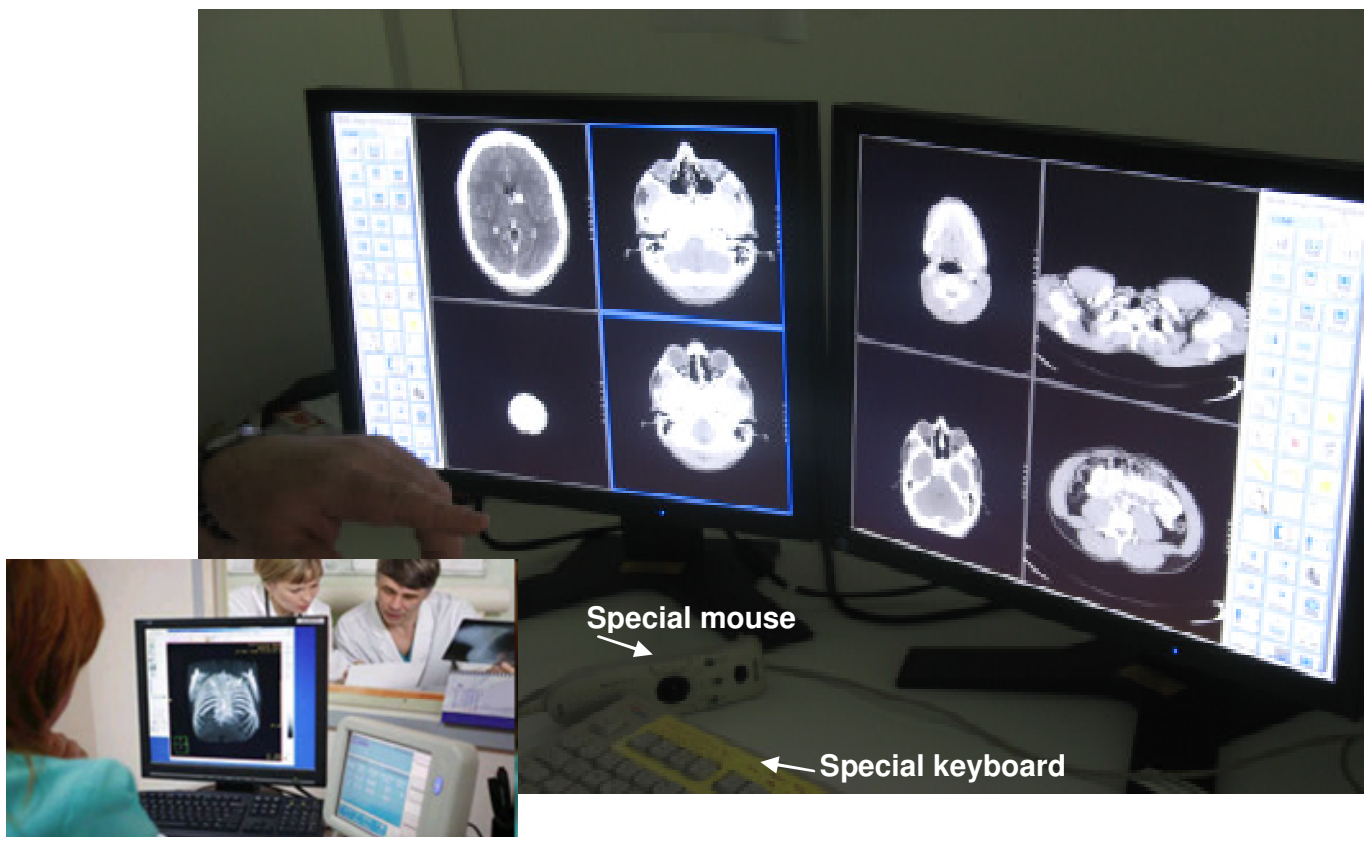

Figure 1. Retrieval and examination of 2D picture series.

The remainder of this book chapter is organized as follows. Section 2 outlines the clinical workflow and interaction requirements. Section 3 describes the knowledge engineering process for image annotations and the dialogue interaction. Section 4 describes the design of multimodal user interfaces and section 4 goes into the implementation of MEDICO's speech-based dialogue shell. In section 5 we discuss and analyze our combined semantic search and retrieval architecture. The final section offers a conclusion and describes our future work in the MEDICO use case.

## 2. Clinical Workflow and Interaction Requirements

To enable the search and understanding of scalable and flexible semantic images, semantic labeling and the interlinking of the data of interest is required. This becomes technically possible when all semantic descriptions are stored in a knowledge base and efficiently linked to previous examinations of the same patient, patient records with a similar diagnosis or treatment, and/or external knowledge resources, such as publications that are relevant in the context of the particular symptoms of the first diagnosis. Several approaches to the semantic annotation of medical images and radiology reports exist. All of these approaches are not only accomplished *offline* but are also quite time-consuming and expensive due to the required user interaction.

We are concerned with answering the following questions:

- How can we enable the semantic annotation of patients' findings without interrupting the clinicians' workflow?
- How can we support the clinical daily tasks in a way that allows parallel semantic annotations of relevant clinical findings without additional efforts?

To address these questions, the following subsection first discuss the today's workflow in radiology. This is followed by a short overview of existing approaches towards the semantic annotation of medical images and radiology findings. This will lead us to the particular requirements for the next generation of radiology workflow supported by semantic and context-sensitive dialogue systems.

## 2.1. Existing Approaches to Semantic Image Annotation

Several approaches to the semantic annotation of medical images and related written findings exist. The approaches differ in the degree of automation and in the underlying data source they start with.

*Automated image parsing* methods, such as those presented in (Seifert et al., 2009) provide means to hierarchically parse whole body CT images and efficiently segment multiple organs while taking contextual information into account. At present, the software is capable of segmenting six organs and detecting 19 body landmarks very quickly and robustly in about 20 seconds. By forming an anatomical network, the landmarks can be used to restrict the search area in the context of organ detection. New anatomy can be easily incorporated since the framework can be trained and handles the segmentation of organs and the detection of landmarks in a unified manner. The detected landmarks and segmented organs are used in multiple ways. First, they facilitate the semantic navigation inside the body (see Figure 2, left), and second, they are used for the generation of semantic annotations such as "spleen" or "splenomegaly".

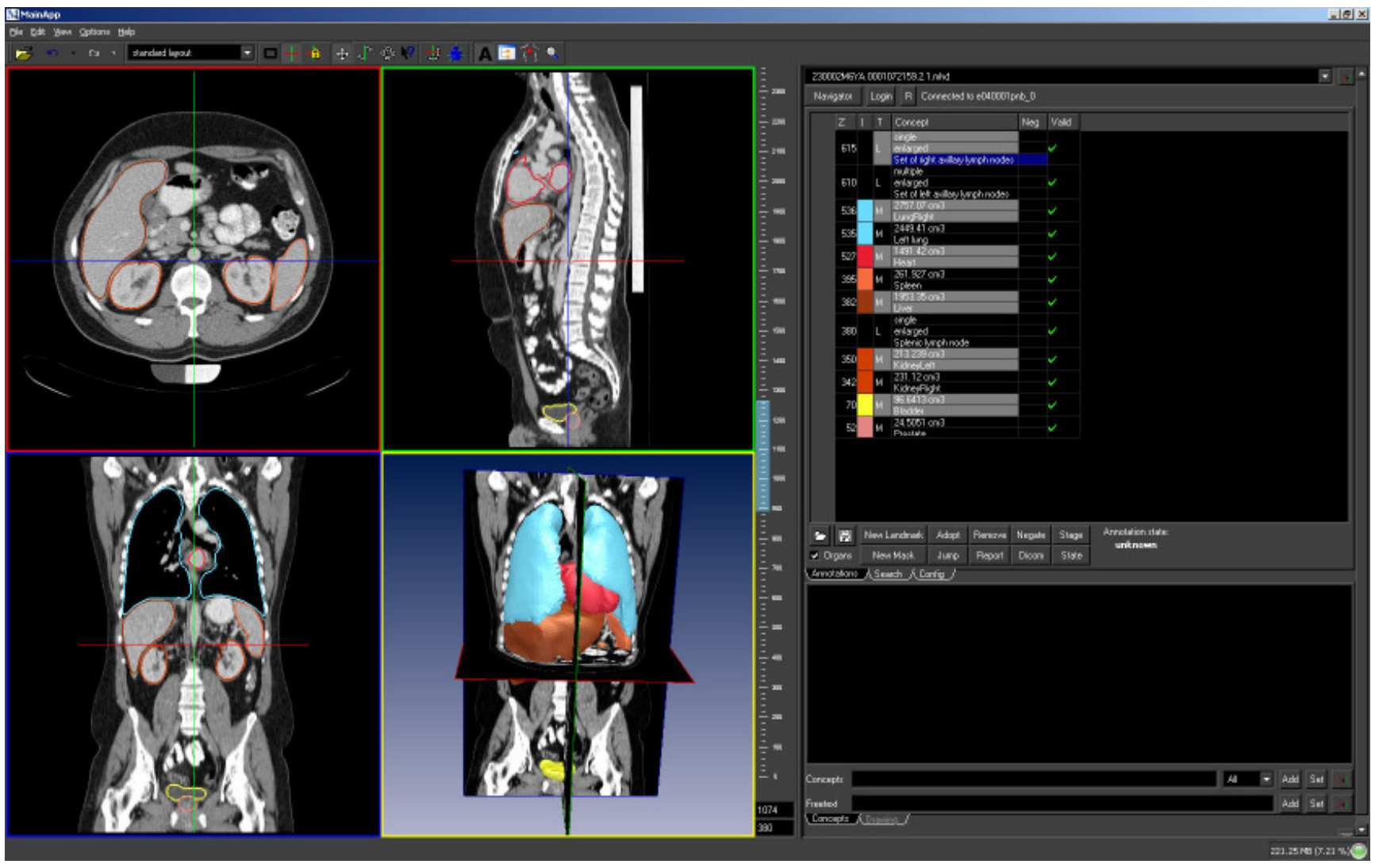

Figure 2. MEDICO application that integrates automatic landmark and organ detection with manual image annotations.

While automated image parsing remains incomplete, *manual image annotation* remains an important complement. MEDICO is only one of several other research projects aiming to integrate manual image annotation in the reporting workflow of radiologists (e.g., the Annotation and Image Markup Project is developing an ontology for medical image annotations, see (Rubin et al., 2008) and (Dameron et al., 2006)). Currently, MEDICO system users can manually add semantic image annotations by selecting or defining anatomical landmarks or arbitrary regions / volumes of interest (see Figure 2, right).

The extraction of information from DICOM headers and DICOM structured reports is another approach to get metadata for semantic image annotation. DICOM (Digital Imaging and Communications in Medicine, http://medical.nema.org/) is the current standardized format used for storing basically all medical images. Metadata such as patient demographics and acquisition parameters are stored in DICOM headers. Within the MEDICO project, we are working towards the automated extracting of DICOM metadata and its conversion into a DICOM ontology, based on OWL which is aligned with our medical image annotation ontology (for details, see (Möller et al., 2009)). With further acceptance of DICOM structured reports, an additional source of semantic image annotations will become available. As described in Part 16 of the DICOM standard, DICOM structured reports are already based on formal clinical healthcare terminology like, e.g., concepts from SNOMED®.

## 2.2. New Radiology Interaction Requirements

The main task in (diagnostic) radiology is to interpret medical images from various modalities like computed tomography or magnetic resonance imaging. Modern radiology information systems automatically route images to the assigned radiologist immediately after the acquisition of the images. Since even a single examination can result in hundreds and even thousands of images, the images are organized according to the DICOM standard into series. A series, for example, contains individual 2D images (“slices”), acquired during one run of a medical imaging device, and these images make up a 3D volume of some body part. Typically, one imaging examination, referred to as a “study” in DICOM, consists of multiple series that are acquired using different machine settings, before or after administration of some contrast media. The series may also contain images from a variety of post-processing options (e.g., to enhance soft tissue contrast).

The process of *reading* the images is highly efficient. While the radiologist views the images in each series essentially in sequential order, he uses a special mouse (Figure 1, below) or keyboard to navigate and manipulate the images (e.g., to zoom, to change display settings, or to perform measurements) while he dictates the image findings that make up his report. Recently, structured reporting was introduced that allows radiologists to use predefined standardized forms for a limited but growing number of specific examinations. However, radiologists feel restricted by these standardized forms and fear a decrease in focus and eye dwell time on the images (Hall, 2009; Weiss et al., 2008). As a result, the acceptance for structured reporting is still low among radiologists while referring physicians and hospital administration in general are supportive of structured standardized reporting since they ease the communication with the radiologists and can be used more easily for further processing (statistics, quality control, alerts, and reminders, etc.).

We strive to overcome the limitations of *structured reporting*:

1. Content-based information should be automatically extracted from medical images.
2. In combination with dialogue-based reporting, radiologists should no longer fill out forms but focus on the images while either dictating the image annotations of the reports to the dialogue system or refining existing annotations.[2]
3. In a further step, individual, speech-based findings should be organized according to a specific body region and structured reports should be generated.

## 2.3. Design and Implementation Strategy

We can identify important design recommendations and usability issues based on the clinical workflow and interaction requirements, with a focus on the new radiology interaction requirements. These recommendations should allow us to implement a multimodal dialogue shell to improve the clinical reporting process, the patient follow-up process, and/or the clinical disease staging and patient management process. Our mission statement "Best medical diagnosis for all" requires the implementation of specialists' contents and interactions from the medical scenario. Furthermore, the design and implementation strategy has to include the integration step into the medical environment. Clinical requirements for a multimodal interface and the integrated multimodal dialogue shell featuring a touchscreen display surface describe the relationship between the "Best medical diagnosis for all" mission statement as a MEDICO requirement and the implementation.

[2] If, for example, he detects a stenosis in a coronary artery, he would simply point to the stenosis, dictate "moderate stenosis", which would be acknowledged by the dialogue system as "moderate stenosis in proximal segment of the right coronary artery". This would make use of the analysis capabilities of MEDICO which allow automatic detection of anatomic locations (Seifert, 2009).

To address the challenges of advanced medical image search while using a dialogue shell, the following four research questions arise:

1) How is the workflow of the clinician, i.e.,
   a) What kind of information is relevant for completion of his daily tasks?
   b) At what stage of the workflow should selected information items be offered?
2) What are the particular challenges and requirements of knowledge engineering in the medical domain?
   a) Can those challenges be addressed by a semi-automatic knowledge extraction process based on clinical user interactions?

In sections 4 and 5, we will describe the multimodal user interface design and implementation stages. With our dialogue shell (we use an upgraded version of the dialogue system for question answering on the Semantic Web developed at DFKI, see (Sonntag et al., 2007b)), we try to smoothly embed the relevant question into the dialogue as initiated by the MEDICO system. The particular requirements for the next generation of radiology workflow should be supported by semantic and context-sensitive dialogue systems.

## 3. KNOWLEDGE ENGINEERING

In our context, we use the term "knowledge engineering" in the sense discussed by (Grüninger and Uschold, 1996). It refers to "methods for creating an ontological and computational basis for reuse of product knowledge across different applications within technical domains." Consequently, we understand ontology management in the medical domain as a specific knowledge engineering task which results in a medical knowledge engineering methodology and the modeling of a domain-specific medical ontology.

Various challenges exist in medical knowledge engineering. One challenge is that the knowledge engineer is not familiar with the complex and comprehensive medical terminology in the medical ontologies. The major challenge, however, is the so-called "knowledge acquisition bottleneck." We cannot easily acquire the necessary medical knowledge that ought to be used in software application but is possessed by medical experts.

To determine the scope and level of detail of the domain's semantics, i.e., the relevant metadata for annotating medical images, the kind of knowledge clinicians are interested in is absolutely relevant. The scope of the constraint domain can be determined by the set of derived query patterns (and dialogue questions), providing guidance in identifying the significant fragments of used ontologies (in our case the Foundational Model of Anatomy (FMA, see Rosse and Mejino, 2003), Radlex (Langlotz, 2006), and ICD-10[3], the International Classification of Diseases). Moreover, the low level features, segmentations, and quantitative measures, derived from automatic image processing, need to be associated with domain ontologies and those ontologies used to retrieve the specific information, e.g., the dialogue ontologies which cover the available interaction forms such as asking questions and providing annotations.

## 3.1. Medical Knowledge Engineering Methodology

From the knowledge engineering requirements, we derived a knowledge engineering methodology that is specific for the medical domain (Wennerberg, 2008). It results in a recommendation study for the three pillars of ontology treatment: knowledge engineering, ontology mediation and alignment (also cf. Noy, 2004), and ontology population and learning (Sonntag et al., 2009). Our ontology engineering approach was constrained by the clinical knowledge requirements upon which we developed the KEMM methodology. To satisfy the radiologist's information need, the semantically integrated image annotations have to be presented to the user in a coherent way. More precisely, the multimodal presentation has to be embedded into a coherent user system natural dialogue. Three typical clinical scenarios which involve the dialogue shell are of interest for further analysis of clinical knowledge requirements:

[3] http://www.who.int/classifications/apps/icd/icd10online

1. The clinical reporting process;
2. The patient's follow-up treatment (i.e., monitoring the patient's health condition and the development of the disease);
3. The clinical disease staging and patient management.

The three clinical scenarios require the acquisition of various types of domain knowledge:

1. The *clinical reporting process* focuses on the general question "What is the disease?" (or, as in the lymphoma case, "which lymphoma?") To answer this question, the *semantic annotations* on medical image contents are used. These are typically anatomical parts such as organs, vessels, lymph nodes, etc.
2. Within the *patient follow-up process,* the clinician's concern is whether or not his former diagnosis hypothesis is confirmed by the outcome of the treatment. In other words, a clinician can only know what he is treating when he sees how the patient responds (Starbucks, 1993).
3. In the *clinical staging* and *patient management process* the general concern is with the next steps in the treatment process. The results of the clinical staging process influence the decisions that concern the patient management process in a later phase.

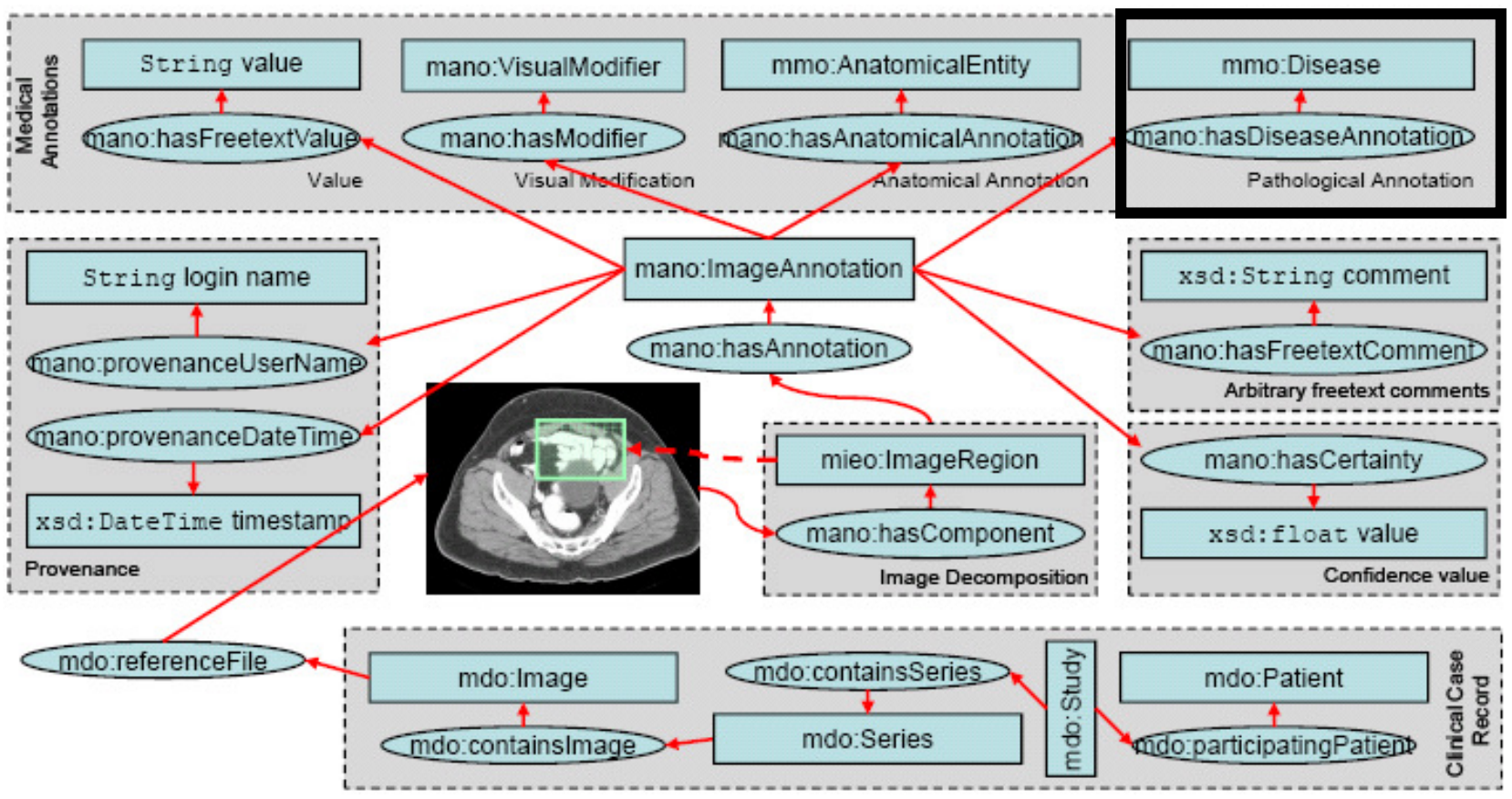

Figure 3. MEDICO semantic annotation scheme.

### 3.2. Ontology Modeling

The system architecture of MEDICO uses a comprehensive and multi-layered ontology. This MEDICO ontology hierarchy is used to represent medical domain knowledge as well as specify the format of image annotations and patient metadata. Using the same representation formalism to represent domain knowledge and annotations allows us to formulate cross-modal and language-independent search queries. During the execution of these queries, the background knowledge from different medical ontologies such as the Foundational Model of Anatomy ontology (FMA), RadLex, and International Classification of Diseases (ICD-10) is used to perform query expansion to retrieve images which are annotated with semantically similar concepts (Figure 3). Further details on the MEDICO ontology hierarchy are covered in (Möller et al., 2009). Our approach to the unification of semantic annotation and querying in biomedical images repositories when using a semantic dialogue shell has been described in (Sonntag and Möller, 2009).

In the context of this book chapter, we will limit ourselves to the modeling of semantic image annotations and the model for storing patient metadata. Figure 3 illustrates the structure (i.e., the schema) of an image annotation. The medical image in the center is decomposed into *ImageRegions*. These are arbitrary segments of medical images or 3D volumes and can be annotated with *ImageAnnotations* in the next step. We differentiate between three dimensions of medical image annotations: (1) for anatomy we use the FMA; (2) the concept for the visual manifestation of an anatomical entity on an image is derived from the modifier and imaging observation characteristic sub-trees of RadLex; (3) we consider the disease dimension as the interpretation of the combination of the previous two. Here we use the ICD-10 as the input source. Additionally, a free text value field can be used to save measurements, e.g., sizes of certain anatomical structures.

Provenance data is stored for the user (currently we use the user's login name) and time stamps are also produced. For automatically acquired image annotations, a respective note is inserted. Additional comments can be saved using the property *hasFreetextComment*. This ensures that annotations which cannot yet be expressed using concepts from the ontology can at least be stored in an informal way and do not get lost.

Additionally, the user can specify a continuous confidence value from the range [0..1] to express his certainty about the actual correctness of each annotation. For automatically acquired image annotations this confidence slot can be used to store the confidence value generated during the feature extraction process to make the accuracy the automatic recognition/extraction process transparent for the medical expert.

The DICOM standard is the most commonly accepted standard to interchange digitized medical images. It provides a container format for data from different modalities such as X-ray, ultrasound, Computed Tomography (CT), etc. Unlike normal photos, e.g., in JPEG format, images in this format contain a broad range of patient and image acquisition metadata in their file headers. The MEDICO ontology also contains its own DICOM ontology which models the hierarchical data structure of the DICOM standard. Essentially, this contains the elements "study" which can contain multiple "series" which, in turn, potentially contains multiple "images." While a study is used to capture all images of a patient acquired for a certain diagnosis, a series collects all images of a single imaging acquisition. The image slices generated by a CT scanner during a single scan are usually grouped into one series.

## 4. Design of Multimodal User Interfaces

Usability applies to every aspect of a research prototype or product with which a person interacts. Every design and development decision made throughout the product cycle has an impact on that product's usability. As (clinicial) users depend more and more on automatic medical software to get their jobs done and use automatic computer systems in more critical use case scenarios (i.e., the *clinical reporting process*), usability can be the critical factor ensuring that the multimodal (dialogue) interface will be successful and used.

### 4.1. Usability Guidelines

Usability guidelines (see, e.g., Garrett, 2002) consider five different planes (Figure 4). Every plane has its own issues that must be considered. From abstract to concrete, these are (1) the strategic plane, (2) the scope plane, (3) the structure plane, (4) the skeleton plane, and (5) the surface plane.

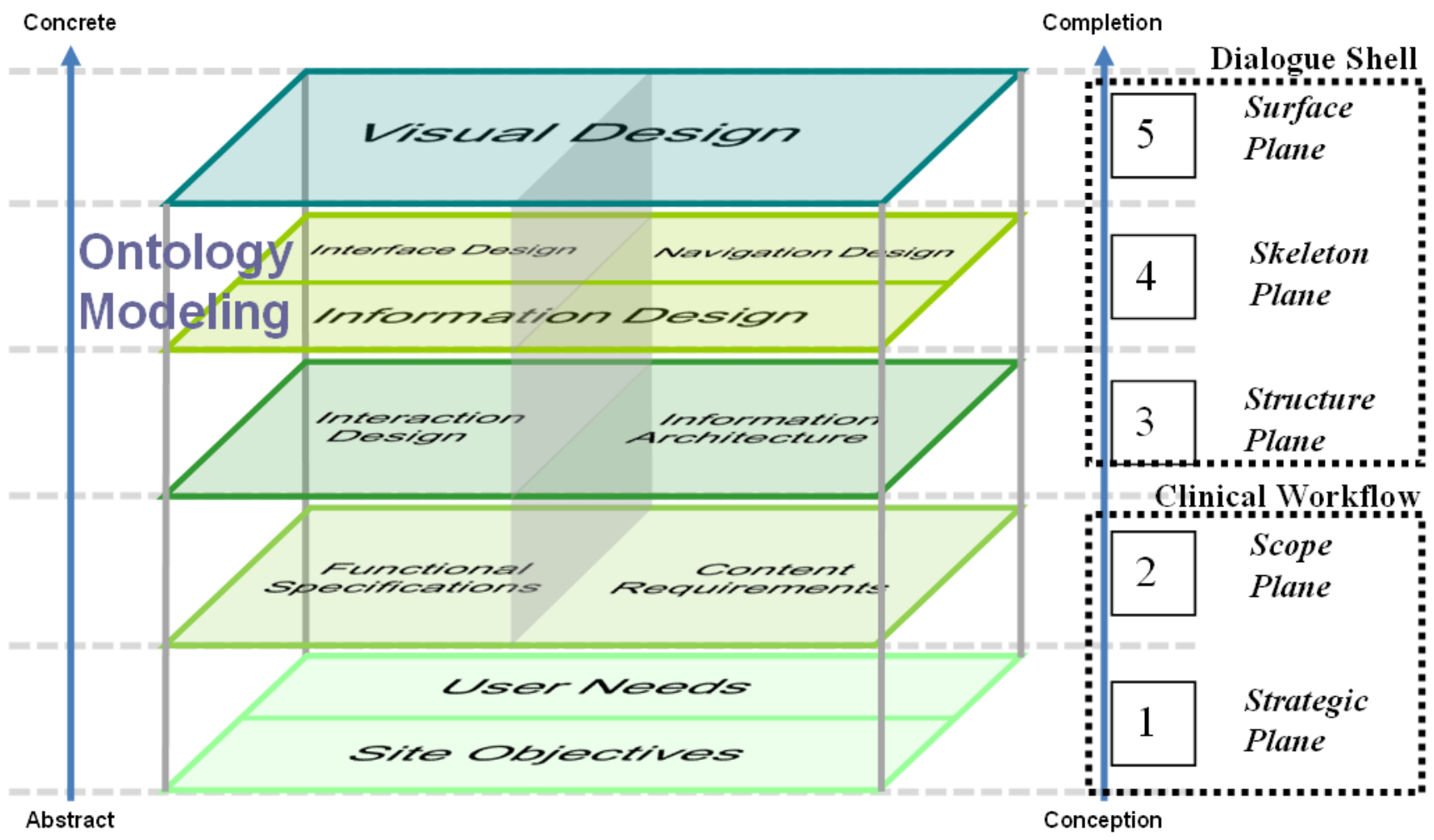

Figure 4. Usability planes and corresponding design issues for implementation.

Defining the users and their needs on the strategic planes is the first step in the design process. It is useful to create personas that represent a special user group. On the scope plane you have to define the system's capacity (cf. *clinical reporting process*) and then the technical requirements. These two planes have already been discussed in section 2 of this chapter as clinical workflow and interaction requirements. The structure, skeleton, and surface planes correspond to the design and implementation of the concrete dialogue shell. The information design of the skeleton plane is represented by the ontologies we modeled in the context of the *clinical reporting process.* This means the skeleton plane is already pre-specified by the ontology engineering requirements in the medical application domain. The design phase for the multimodal user interface (i.e., the dialogue shell) is restricted to the interaction design/information architecture storyboard on the structure plane and the speech and touchscreen interaction design on the surface plane (described in more detail).

## 4.2. Storyboard (Structure Plane)

The design task for the structure plane consists of a cycle of action and reaction. Either the user acts and the system reacts or the other way around. Every time the user uses the dialogue system, she will improve her mental model of the system. But this only works if the conceptual model of the system matches the user's mental model. If the user can predict what the system will do, she is more willing to do trial and error. For this purpose, a storyboard is constructed and implemented by concrete SIEs (Semantic Interface Elements, see Sonntag et al., 2009). Figure 5 shows the interaction storyboard and the included SIEs, i.e., Image Annotation SIE (1), Patient Finding SIE (2), Patient Search SIE (3), Browser SIE (4), and Video SIE (5). The touchscreen background SIE is displayed in (B). These SIEs represent the visual interaction elements for MEDICO patient images and patient records. The implementation of the dialogical interaction sequences in the dialogue shell, and the reference dialogue, are based on these visual elements.

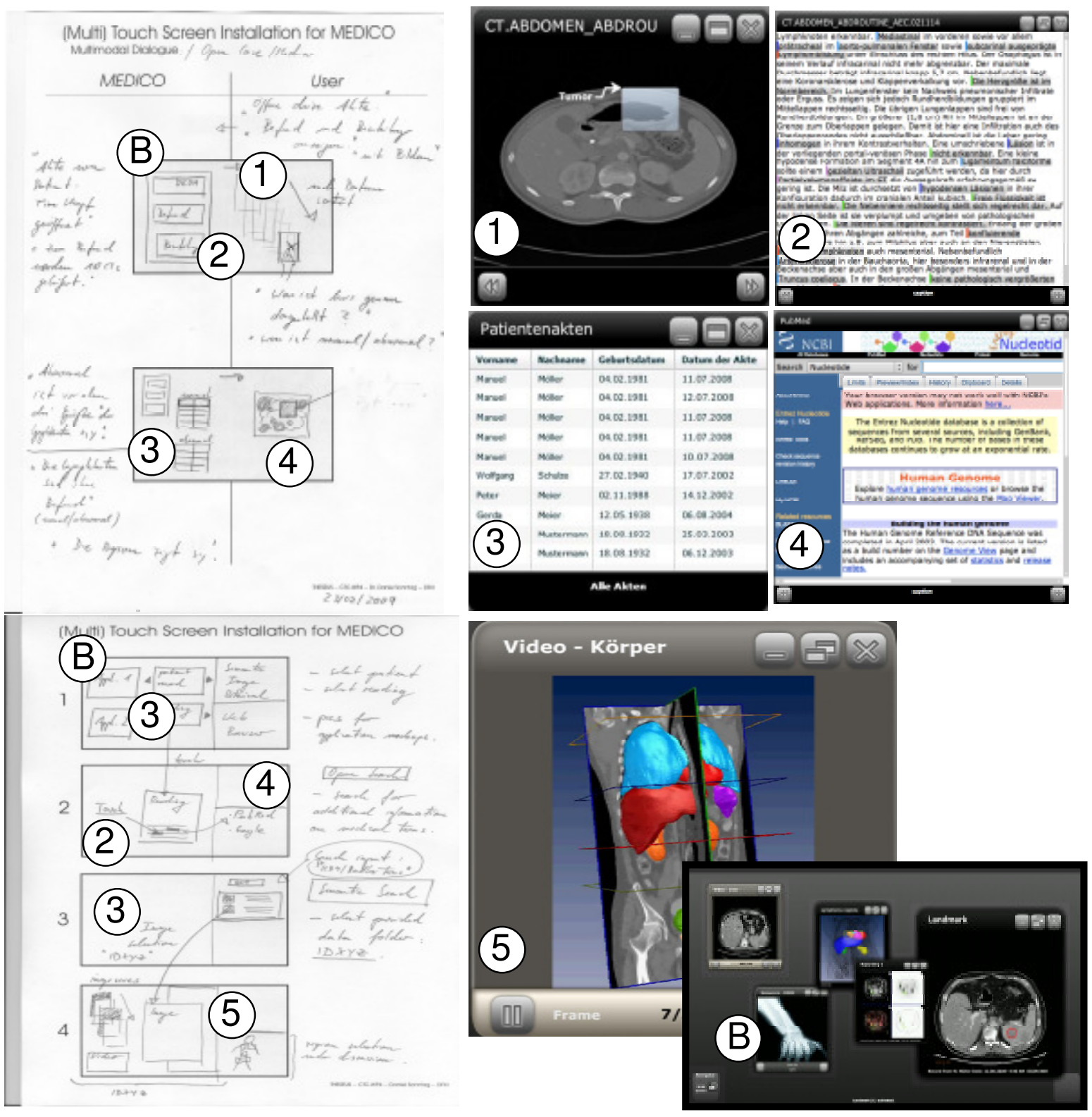

Figure 5. (Left) Interaction storyboard. (Right) Medical semantic interface elements (SIEs), included in the touchscreen installation.

### 4.3. Speech and Touchscreen Interaction Design (Surface plane)

This plane deals with the logical arrangements of the design elements. In the case of a multimodal dialogue system, the logical arrangement results in a user-system natural dialogue whereby the user input is speech and touch and the system output is generated speech or the generation of SIEs which display windows for images, image regions, or other supported interaction elements. The implemented clinical workflow is best explained by example. Consider a radiologist (R) at his daily work of the *clinical reporting process* (also cf. section 3.1) with the speech-based semantic dialogue shell (S):

| | |
|---|---|
| 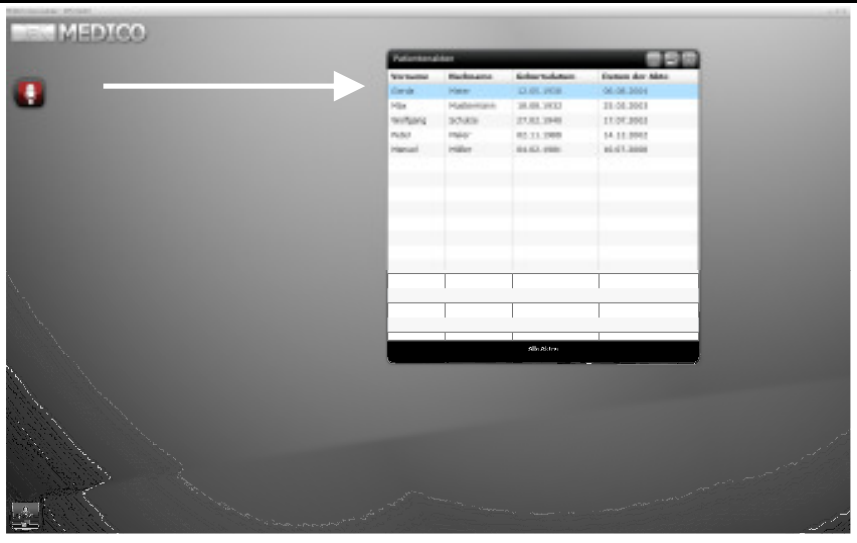  | The potential application scenario (provided by Siemens AG) includes a radiologist which treats a lymphoma patient; the patient visits the doctor after chemotherapy for a follow-up CT examination.<br>**R:** “Show me my patient records, lymphoma cases, for this week.”<br>**S:** Shows corresponding patient records. |
| 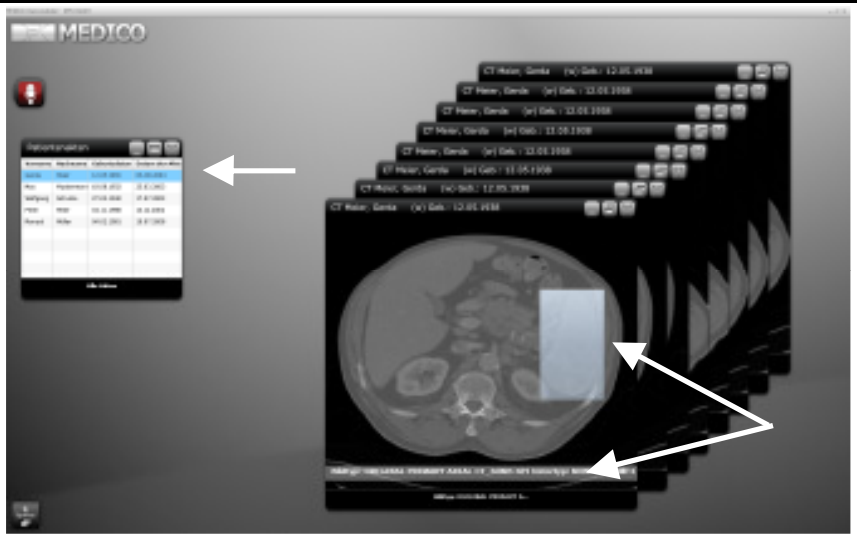  | **R:** “Open the images, internal organs: lungs, liver, then spleen and colon of this patient (+ pointing gesture (arrow)).”**S:** Shows corresponding patient image data according to referral record.<br>The presentation planer of the dialogue system rearranges the semantic interface elements (SIEs). The top-most picture frame, showing the patient information in the header, is interactive; when touching it, special image regions and region annotations are highlighted (two arrows).<br>**R:** Switches to the 5th image and clicks on a specific region (automatically determined). |
| 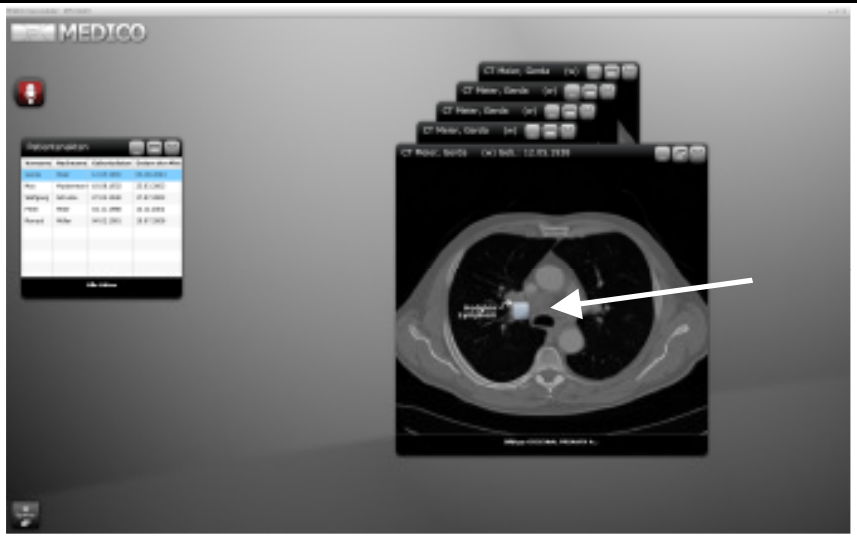  | **S:** The system rearranges the semantic interface elements (SIEs) to signalize that the dialogue focus is on regions.<br>**R:** "This lymph node here (+ pointing gesture), annotate Hodgkin-Lymphoma."<br>**S:** Annotates the image with RDF annotations (cf. Figure 3, highlighted pathological part) and displays a label for the recognized ICD-10 term.<br>**R:** "Find similar lesions with characteristics: hyper-intense and/or coarse texture." |
| 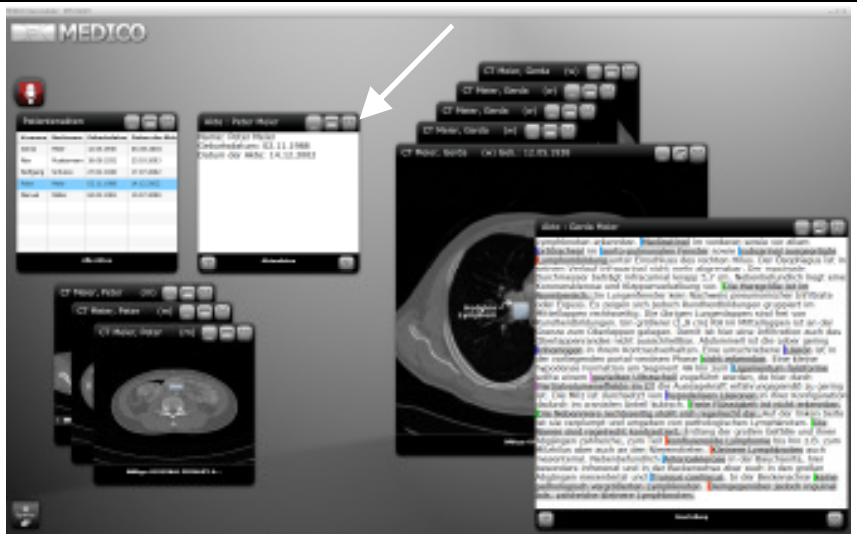  | **S:** MEDICO displays the search results in the record table (also see first screenshot) ranked by the similarity and match of the medical terms that constrain the semantic search (left) and opens the first hit, Peter Maier (arrow), the record, and his images that correspond to the search. The system rearranges the SIEs for the two patients for a comparison.<br>**R:** "Get the findings of this patient"<br>**S:** Opens the findings (text) and highlights the medical terms in different groups. |

One of the radiologist's goals is to estimate the effectiveness of the administered medicine. In order to finish the reading / pathology, additional cases have to be taken into account for comparison. We try to find these cases by matching the medical RDF annotations (FMA, RadLex, ICD-10) of different patient cases stored in the patient triple store. Semantic interface elements allow for a user-friendly interaction with retrieved data presented on the screen, according to the guiding principle "no presentation without representation" (Maybury and Wahlster, 1998). These objects together with their underlying ontology-based representation can then be referenced by the user in the subsequent speech input.

The prototypical installation is a large-screen multimodal interface in which two aspects are implemented: (1) the annotation of radiological images by use of speech and gestures, and (2) the inspection of and navigation through the patients' data. This allows the radiologist to easily come to a diagnostic analysis of the images. The underlying dialogue system makes use of ontology-based retrieval and annotation and, furthermore, enables access to semantic web services in the medical domain.

Additional storyboards have been developed, e.g., in the context of matching different terminologies (ontology matching) for anatomical parts (Sonntag, 2008). All storyboards have been prototypically implemented while using our speech-based dialogue shell.

## 5. Implementation of the Speech-based Dialogue Shell

Within a multimodal dialogue system two or more user input modes, such as speech, gestures and other input modalities are proceed in a coordinated manner. The various input modalities can be combined. Our multimodal dialogue system is based on the Ontology-Based Dialogue Platform, ODP, which provides a lightweight open architecture for the flexible integration of multimodal dialogue processing components (Wahlster 2003; Wahlster 2006).

A generic architecture of a multimodal dialogue system is illustrated in Figure 6. It consists of components for the following tasks:

- Recognition of multimodal input, e.g., automatic speech recognition;
- The interpretation of the multimodal input including modality fusion;
- The dialogue and interaction management for the system behavior;
- The semantic access to the backend application and services, including interactive semantic mediation and semantic mashups (also see Figure 9);
- The presentation planning and realization;
- And the fission of the output modalities.

Input and output components can be attached to the generic system. Such components include a speech recognizer (*ASR*) and a speech synthesis (*TTS*) module. Our approach relies on a flexible toolbox of generic and configurable dialogue shell building blocks. The exchange data between the different modules implemented upon the mentioned building blocks is based on ontology-based data using so called *extended Type Feature Structures* (eTFS) (Pfleger and Schehl, 2006).

Besides the presented use case related to the medical domain, the ODP framework (an ontology-based dialogue platform available at http://www.semvox.de/) has been used to build prototype systems for various application scenarios. TEXO Mobile (Porta et al., 2009), developed within the THESEUS research program, provides a mobile, multi-modal interface for accessing business web services. A further application is the CoMET system. CoMET (Collaborative Media Exchange Table) provides speech-enabled semantic access to personal multimedia content and related online services for music-oriented entertainment. It demonstrates how users intuitively exchange information and media using spoken language and gestures; photo, video, and music files can be grouped, annotated, shared, or simply played back.

## 6. Combined Semantic Search and Retrieval Architecture

To make the results of the automatic object recognition algorithms available for semantic search, we had to integrate disparate techniques into a hybrid system. The automatic object recognition performs an abstraction process from simple low-level features to concepts represented in formal ontologies. For performance reasons, medical image processing libraries are almost exclusively implemented using C and C++. At the same time, libraries for handling data in the Semantic Web standards OWL and RDF are most advanced in Java.

Figure 7 shows the overall architecture of our approach for integrating manual and automatic image annotation. One of the main challenges was to integrate the C++ code for object recognition (left) with the MITK-based image viewer, the annotation tool (bottom) also in C++, and the Java-based components for knowledge base manipulation and semantic search (right). We came up with a distributed architecture with a CORBA (Common Object Requesting Broker Architecture) server as a mediator between our C++ and Java components.

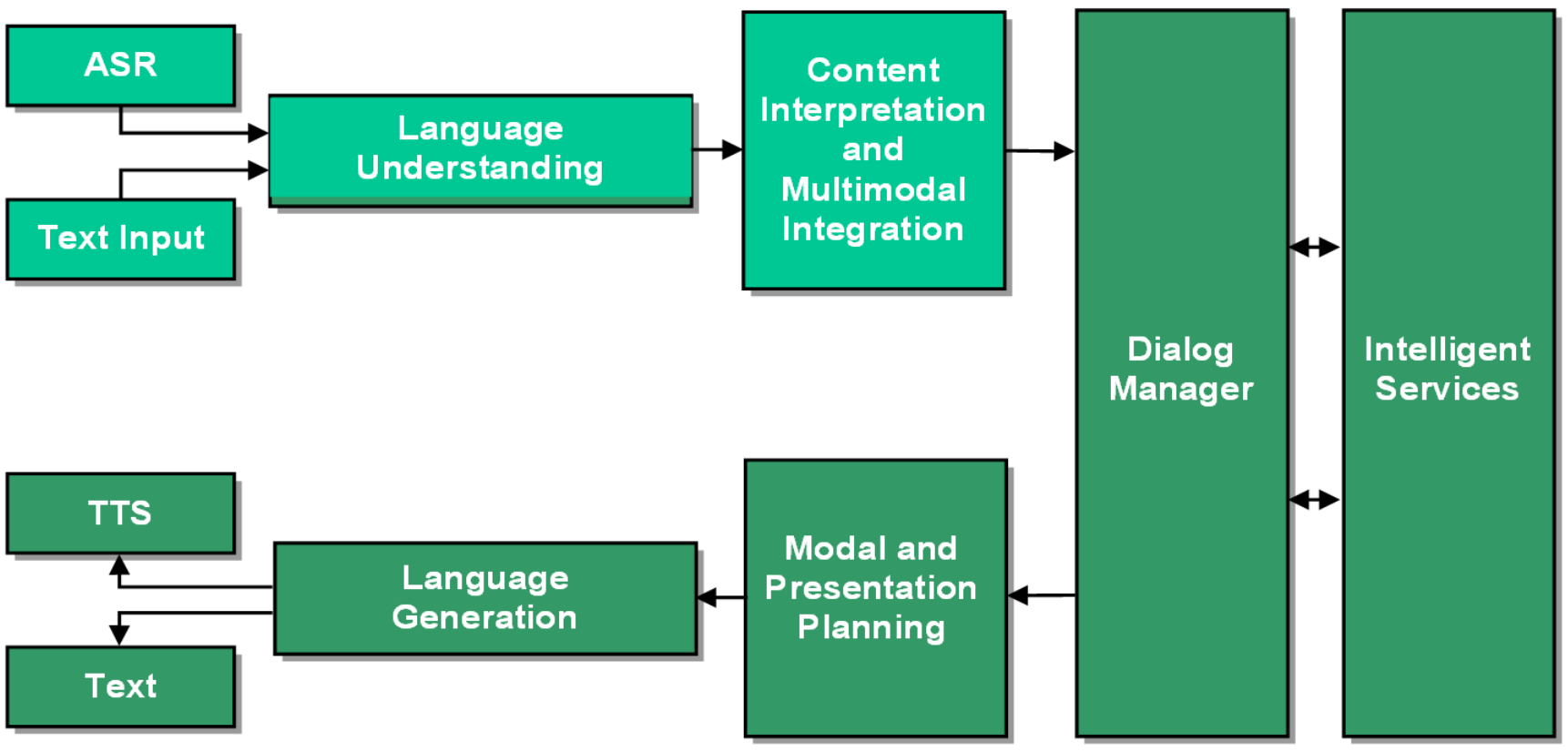

Figure 6. Generic Architecture of a Multimodal Dialogue System.

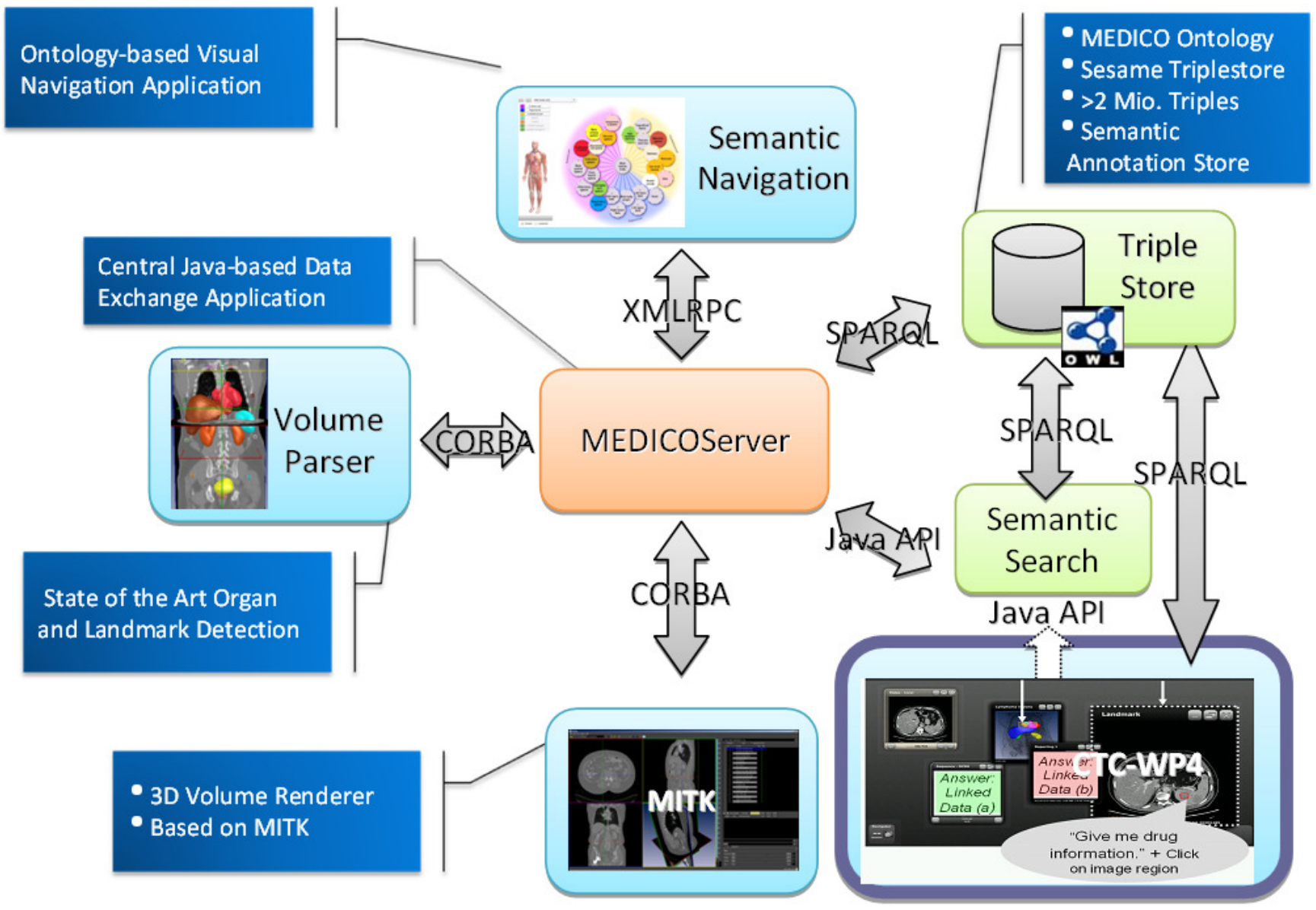

Figure 7. Overall MEDICO Semantic Search Architecture.

## 6.1. Medico Server and Components

Instances of the automatic object recognition system (potentially distributed across different machines at remote locations) can register with the central CORBA server. From the automatic object recognition system, all detected landmarks are sent together with a volume data set identifier to the CORBA server. To identify volume data sets, we use the Study and Series Instance UID as defined in the DICOM standard.

***Volume parser***

For automatic object recognition we use a state-of-the-art anatomical landmark detection system described in (Seifert, 2009). It uses a network of 1D and 3D landmarks and is trained to quickly parse 3D CT volume data sets and estimate which organs and landmarks are present as well as their most probable locations and boundaries. Using this approach, the segmentation of seven organs and detection of 19 body landmarks can be obtained in about 20 seconds with state-of-the-art accuracy below 3 mm mean mesh error and has been validated on 80 CT full or partial body scans (Seifert, 2009).

***Triple store***

For the central Triple Store we chose Sesame (Broekstra, 2001) because of its easy online deployment and fast built-in persistence strategy. Deployed to a central application server, Sesame provides the system with a central RDF repository for storage and retrieval of information about the medical domain, clinical practice, patient metadata, and image annotations. This central repository offers different interfaces for data retrieval and manipulation. They provide access to two different abstraction layers of the data. On the low level, a direct access to the RDF statements is possible using the query language SPARQL (Prud'hommeaux and Seaborne, 2007). The semantic dialogue shell directly accesses the Triple Store via SPARQL commands in order to retrieve patient images with semantic annotations.

***Semantic search***

More complex functions such as query expansion based on the hierarchical information in the ontologies as well as all data manipulation operations are preformed through custom API libraries. The manual annotation and semantic search application uses the same RDF repository for data storage and retrieval and thus has direct access to the automatic annotation results.

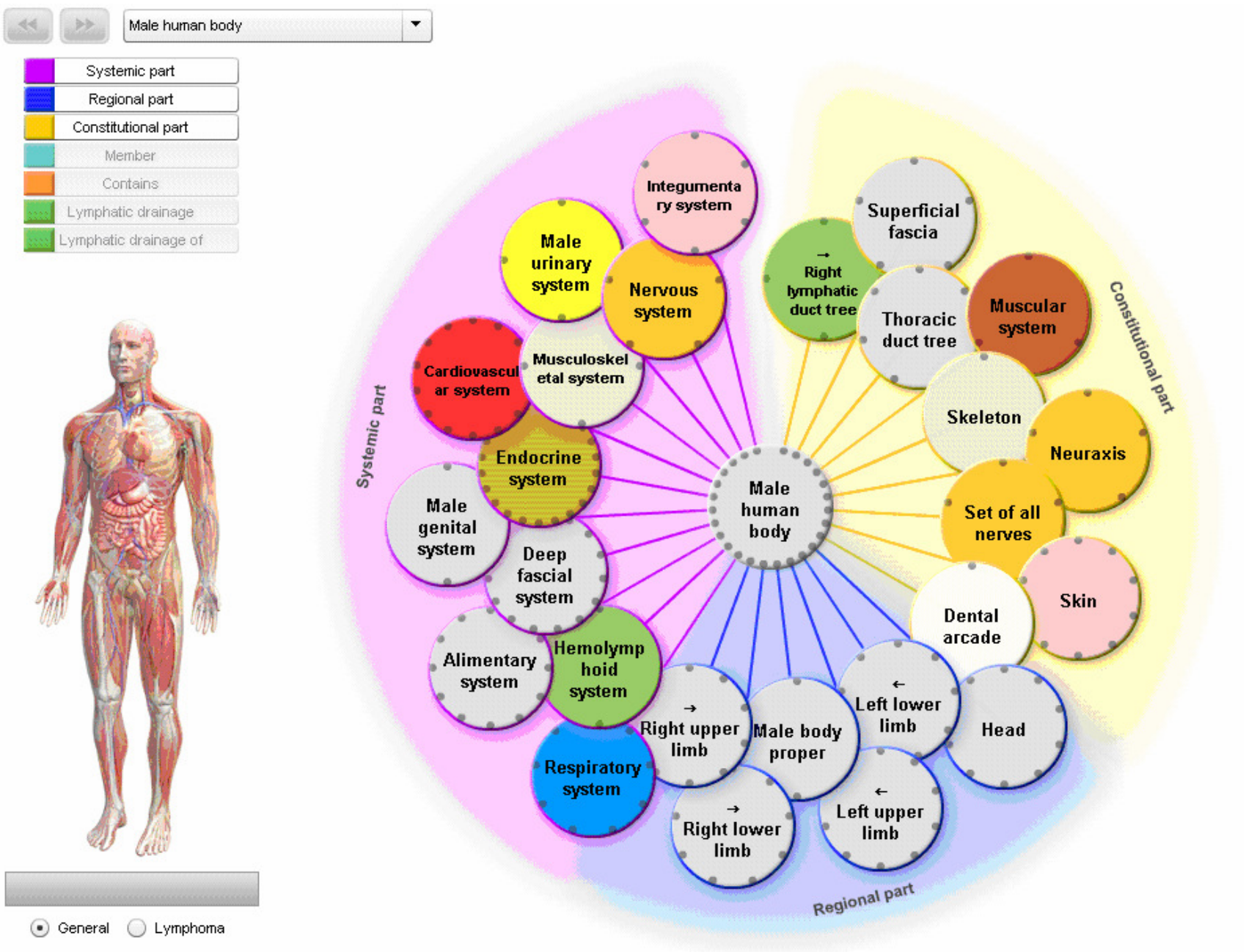

Figure 8. Semantic Navigation Interface Element.

***Semantic navigation***

Semantic Navigation shows anatomical concepts in a browser window. This window can be accessed by the dialogue shell through the XML RCP / Java Interface. In this way, additional *clinical reporting process* relevant information can be accessed by the radiologist (Figure 8).

## 6.2. Search Architecture of the Multimodal Dialogue Shell

The technical semantic search architecture of the multimodal dialogue shell (cf. CTC-WP4 in Figure 7) comprises of three tiers: the application layer (user interface, dialogue system/manager), the query model/semantic search layer (eTFS/SPARQL structures), and the dynamic knowledge bases layer for the application backend (Figure 9). The intelligent services (cf. Figure 6) are represented by the medical information sources in our dynamic knowledge base layer (Figure 9, right).

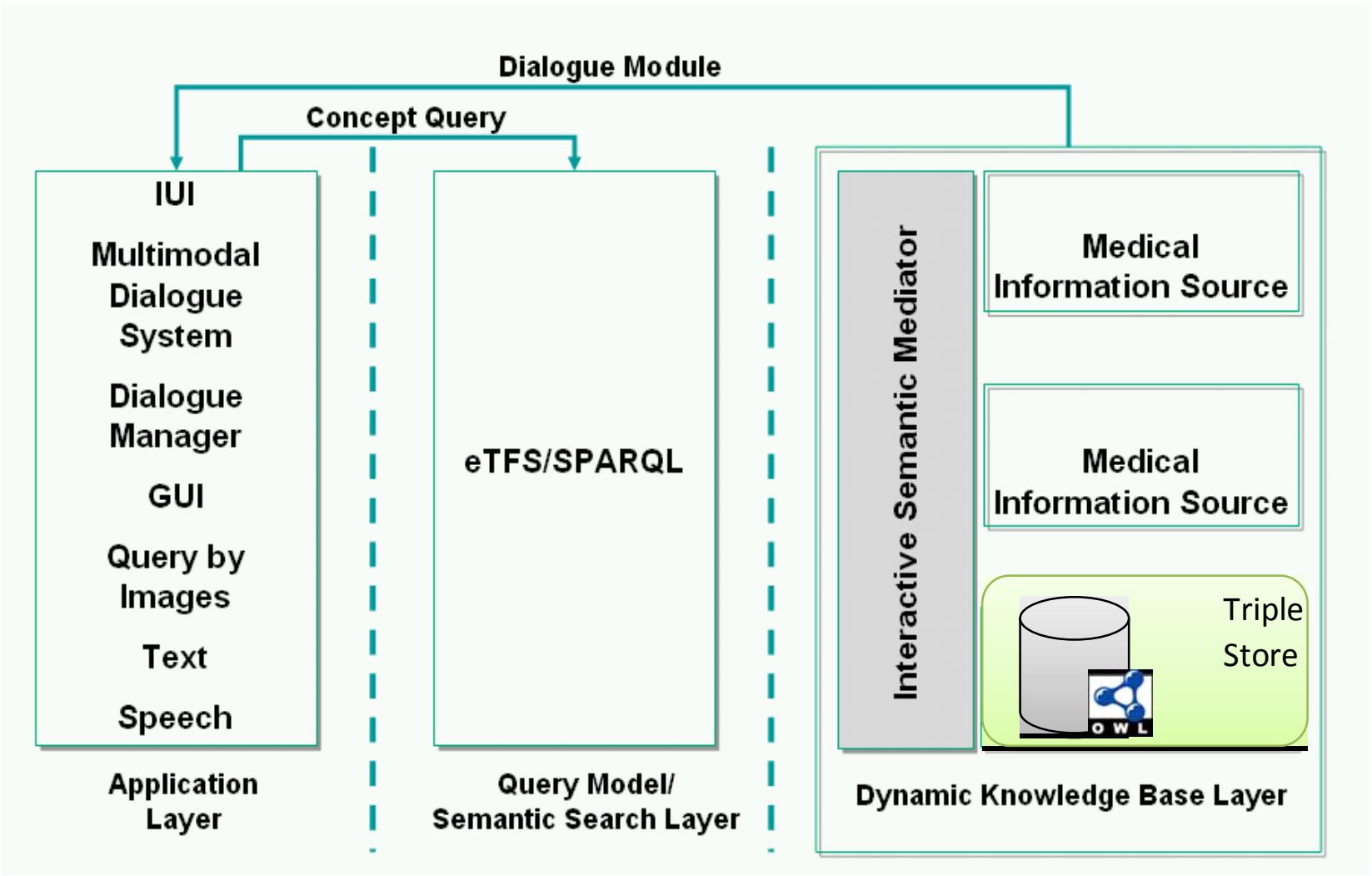

Figure 9. Three Tier Search Architecture.

In the context of this contribution, we will focus on the dynamic knowledge base layer. An interactive semantic mediator component is responsible for providing an integrated view of the data. There are three structurally different medical information sources, i.e., the Triple Store accessed via SPARQL queries, the Semantic Search functionality accessed via a Java API and the Semantic Navigation application accessed via a Java wrapper API. In this situation, very different semantic resources have to be matched at the query or answer side while using the dialogue-based image retrieval functionality. The Triple Sore, however, contains the most important data, the image region annotations. While using the dialogue-based annotation functionality (also cf. the multimodal dialogue in section 4.3), we also access the Triple Store for storing the speech-based image (region) annotations. The semantic mediator provides the necessary transformations especially between the structurally different data sources. In future work, medical Linked Data sources, e.g., LODD, (normally at SPARQL endpoints) will be addressed by the help of this component, too. The same applies to the connection between the dialogue shell and the MITK visualization tool.

## 7. Conclusion and Future Work

We described the design and implementation of a semantic dialogue system for radiologists in a comprehensive case study. In intensive discussions with clinicians we analyzed how the use of semantic technologies can support the clinician's daily work tasks, apart from the fact that in daily hospital work, clinicians can only manually search for *similar* images—for which we provided a solution, the semantic dialogue shell for radiologists. We discussed the clinical workflow and interaction requirements and focused on the design and implementation of the multimodal user interface for image search and image region annotation and its implementation while using a speech-based dialogue shell.

The overall MEDICO Semantic Search Architecture which includes our CTC-WP4 semantic dialogue shell will now be tested in a clinical environment (University Hospitals Erlangen). Furthermore, the question of how to integrate semantic image knowledge with other types of data, such as patient data, is paramount. For clinical staging and patient management the major concern is which procedure step has to be performed next in the treatment process.

A completely *new* approach for including text semantics seeks for the *semi-automatic extraction of terms and relations in radiology reports* as generated by clinicians in the process of analyzing the patient's findings by studying medical imaging data. Radiology reports are dictated documents, and although they are stored as written documents, they are only seldom written in complete sentences and grammatical constructions. For instance, many sentences lack verbs and punctuations. In addition, abbreviations are very common and temporal and spatial information for describing image content is used extensively. Due to those textual particularities, existing approaches for natural language analysis (Hirst and Budanitsky, 2006) need to be customized and new methods need to be developed. Ongoing work in the MEDICO project has the goal to develop means for automatic knowledge extraction from radiology reports. In the future, the semantic dialogue shell should display the structured patient reports and allow a radiologist to refer to text passages while using the speech-based system.

## ACKNOWLEDGEMENTS

This research has been supported in part by the THESEUS Program which is funded by the German Federal Ministry of Economics and Technology under the grant number 01MQ07016. We would like to thank Matthieu Deru for the implementation of the semantic interface elements. Our thanks go also out to the semantic dialogue shell team at DFKI and the MEDICO use case, in particular Norbert Reithinger, Robert Neßelrath, Daniel Porta, Gerhard Sonnenberg, Gerd Herzog, Malte Kiesel, Simon Bergweiler, Anselm Blocher, Tilman Becker, Michael Sintek, Pinar Wennerberg, and Colette Weihrauch.

## REFERENCES

(Broekstra, 2001) Broekstra, J. & Kampman, A. (2001). Sesame: A Generic Architecture for Storing and Querying RDF and RDF Schema. In: *Administrator*, Nederland b.v.

(Dameron et al., 2006) Dameron, O., Roques, E., Rubin, D., Marquet, G. & Burgun, A. (2006). *Grading lung tumors using OWL-DL based reasoning*. In: Proceedings of 9th International Protégé Conference.

(Garrett, 2002) Garrett, J. J. (2002). The Elements of User Experience. In: *American Institute of Graphic Arts*, New York, USA.

(Grüniger and Uschold, 1996) Grüninger, M. & Uschold, M. (1996). Ontologies: Principles, methods and applications. In: *Knowledge Engineering Review*, *1(2)*, 93-155.

(Hall, 2009) Hall, Ferris, M. (May 2009). The Radiology Report of the Future. In: *Radiology*, *Volume 251*, Number 2.

(Hirst and Budanitsky, 2006) Hirst, A. & Budanitsky, G. (2006). Evaluating Wordnet-based measures of lexical semantic relatedness. *Computational Linguistics*, *32(2)*, 13-47. Cambridge, PA: MIT Press.

(Langlotz, 2006) Langlotz, C. P. (2006). Radlex: A new method for indexing online educational materials. In: *RadioGraphics*, *26*, 1595-1597.

(Maybury und Wahlster, 1998) Maybury, M. & Wahlster, W. (Eds.) (1998). Readings in Intelligent User Interfaces. *Morgan Kaufmann Publishers*, Inc.

(Möller et al., 2009) Möller, M., Regel, S. & Sintek, M. (2009). RadSem: *Semantic Annotation and Retrieval for Medical Images*. In: Proc. of The 6th Annual European Semantic Web Conference (ESWC).

(Noy, 2004) Noy, N. (2004). Tools for mapping and merging ontologies. In: S. Staab, & R. Studer, (Eds.). *Handbook on Ontologies*, PA: Springer-Verlag (365-384).

(Pfleger and Schehl, 2006) Pfleger, N. & Schehl, J. (2006). Development of advanced dialog systems with PATE. In Proc. of INTERSPEECH 2006—ICSLP: *Ninth International Conference on Spoken Language Processing*, Pittsburgh, PA, USA, pages, 1778-1781, Pittsburgh, PA

(Porta, et al., 2009) Porta, D., Sonntag, D. & Neßelrath, R. (2009). New Business To Business Interaction: Shake your iPhone and speak to it. In: *Proceedings of the11th International Conference on Human Computer Interaction with Mobile Devices and Services*, (MobileHCI)

(Prud'hommeaux and Seaborne, 2007) Prud'hommeaux, E. & Seaborne, A. (2007). SPARQL Query Language for RDF, W3C.

(Rosse and Mejino, 2003). Rosse C., & Mejino, J.L. (2003). A reference ontology for bioinformatics: the foundational model of anatomy. *Journal of Biomedical Informatics*, *(36)*, 478-500.

(Rubin et al., 2008) Rubin, D., Mongkolwat, P., Kleper, V., Supekar, K. & Channin, D. (2008). Medical imaging on the semantic web: Annotation and image markup. In: AAAI Spring Symposium Series, *Semantic Scientific Knowledge Integration*, Stanford, USA

(Seifert et al., 2009) Seifert, S., Barbu, A., Zhou, S., Liu, D., Feulner, J., Huber, M., Suehling, M., Cavallaro, A. & Comaniciu, D. (2009). *Hierarchical parsing and semantic navigation of full body CT data*. In: SPIE Medical Imaging.

(Sonntag, 2007a) Sonntag, D. (2007a). Embedded Distributed Text Mining and Semantic Web Technology. In: Proceedings of the NATO Advanced *Study Institute Workshop on Mining Massive Data Sets for Security*, PA: NATO Publishing.

(Sonntag, 2007b) Sonntag, D., Engel, R., Herzog, G., Pfalzgraf, A, Pfleger, N., Romanelli, M. & Reithinger, N. (2007). SmartWeb Handheld. Multimodal interaction with ontological knowledge bases and semantic web services (extended version). In T., Huang, A., Nijholt, M. Pantic, & A. Plentland, (Eds.). *LNAI Special Volume on Human Computing*, *Vol. 4451*, Berlin, Heidelberg, PA: Springer Verlag.

(Sonntag, 2008) Sonntag, D. (2008). Towards Dialogue-Based Interactive Semantic Mediation in the Medical Domain. In: *Proceedings of the Third International Workshop on Ontology Matching*, (OM-2008) collocated with the 7th International Semantic Web Conference (ISWC).

(Sonntag et al., 2009) Daniel Sonntag, Matthieu Deru and Simon Bergweiler (2009). Design and Implementation of Combined Mobile and Touchscreen-Based Multimodal Web 3.0 Interfaces. *Proceedings of the 2009 International Conference on Artificial Intelligence* (ICAI)

(Sonntag and Möller, 2009) Sonntag, D. & Möller, M. (2009). Unifying Semantic Annotation and Querying in Biomedical Images Repositories. In: *Proceedings of the First International Conference on Knowledge Management and Information Sharing (KMIS)*, IC3K

(Smeulders et al., 2000) Smeulders, A. W. M., Worring, M., Santini, S., Gupta, A. & Jain, R. (2000). Content-based image retrieval at the end of the early years. In: *IEEE Transactions on Pattern Analysis and Machine Intelligence*, *22 No*, 12 1349-1380.

(Weiss et al., 2008) Weiss, D. L. & Langlotz, C. P. (Dec. 2008). Structured Reporting: Patient Care Enhancement or Productivity Nightmare? *Radiology*, *Volume 249*, Number 3.

(Wahlster, 2003) Wahlster, W. (2003). Towards Symmetric Multimodality: Fusion and Fission of Speech, Gesture, and Facial Expression. In: Andreas Günter, Rudolf Kruse and Bernd Neumann (Eds.). KI 2003: Advances in Artificial Intelligence. *Proceedings of the 26th German Conference on Artificial Intelligence*, September 2003, Hamburg, Germany (pg 1-18) Berlin, Heidelberg, Springer, LNAI 2821.

(Wahlster, 2006) W. Wahlster, (Ed.) (2006). SmartKom: *Foundations of Multimodal Dialogue Systems*, Springer, Berlin.

(Wennerberg et al., 2008) Wennerberg, P., Zillner, S., Moeller, M., Buitelaar, P. & Sintek, M. (2008). KEMM: A knowledge engineering methodology in the medical domain. In: C. Eschenbach, & M. Grüninger, (Eds.). Proceedings 5th international conference on formal ontology in information Systems (FOIS). PA: IOS Press.